\documentclass[runningheads]{llncs}
\usepackage{graphicx}
\usepackage{tabularx}  
\usepackage{amsmath}
\usepackage{multirow,multicol} 
\usepackage{setspace} 
\usepackage{floatrow}
\newfloatcommand{capbtabbox}{table}[][\FBwidth]
\usepackage{amsfonts} 
\usepackage{subcaption}
\usepackage[misc]{ifsym} 

\begin{document}

%
\title{Voxel-wise Cross-Volume Representation Learning for 3D Neuron Reconstruction}
%
%
\author{Heng Wang$^{1}$, Chaoyi Zhang$^{1}$, Jianhui Yu$^{1}$, Yang Song$^{2}$, Siqi Liu$^{3}$, Wojciech Chrzanowski$^{4,5}$, Weidong Cai$^{1 (\textrm{\Letter})}$}

\authorrunning{H. Wang et al.}
\titlerunning{VCV-RL for 3D Neuron Reconstruction}
%
\institute{$^{1}$ School of Computer Science, University of Sydney, Sydney, Australia\\ \email{tom.cai@sydney.edu.au}\\
$^{2}$ School of Computer Science and Engineering, University of New South Wales, Sydney, Australia \\
$^{3}$ Paige AI, New York, NY, USA\\
$^{4}$ Sydney Pharmacy School, University of Sydney, Sydney, Australia\\
$^{5}$ Sydney Nano Institute, University of Sydney, Sydney, Australia}

\maketitle              
\begin{abstract}
Automatic 3D neuron reconstruction is critical for analysing the morphology and functionality of neurons in brain circuit activities. However, the performance of existing tracing algorithms is hinged by the low image quality. Recently, a series of deep learning based segmentation methods have been proposed to improve the quality of raw 3D optical image stacks by removing noises and restoring neuronal structures from low-contrast background. Due to the variety of neuron morphology and the lack of large neuron datasets, most of current neuron segmentation models rely on introducing complex and specially-designed submodules to a base architecture with the aim of encoding better feature representations. Though successful, extra burden would be put on computation during inference. Therefore, rather than modifying the base network, we shift our focus to the dataset itself. The encoder-decoder backbone used in most neuron segmentation models attends only intra-volume voxel points to learn structural features of neurons but neglect the shared intrinsic semantic features of voxels belonging to the same category among different volumes, which is also important for expressive representation learning. Hence, to better utilise the scarce dataset, we propose to explicitly exploit such intrinsic features of voxels through a novel voxel-level cross-volume representation learning paradigm on the basis of an encoder-decoder segmentation model. Our method introduces no extra cost during inference. Evaluated on 42 3D neuron images from BigNeuron project, our proposed method is demonstrated to improve the learning ability of the original segmentation model and further enhancing the reconstruction performance.

\end{abstract}
\section{Introduction}
3D neuron reconstruction is essential for analysis of brain circuit activities to understand how human brain works~\cite{app2,liu2016rivulet,liu2018automated,wang2018memory}. 
It traces neurons and reconstructs their morphology from 3D light microscopy image stacks for neuroscientists to investigate the identity and functionality of neurons. 
Traditional tracing algorithms rely on hand-crafted features to capture neuronal structures but they are sensitive to the image quality. 
However, due to various imaging conditions, obtained neuron images suffer from different extent of noises and uneven labelling distribution. 
To attain better tracing performance for 3D neuron reconstruction, an accurate segmentation method to distinguish a neuron voxel from its surrounding low-contrast background is in high demand and necessary. 
However, due to the complexity of neuronal structures and various imaging artefacts, the precise restoration of neuronal voxels remains a challenging task.

A line of deep learning based segmentation models~\cite{hanchuan,jiezhao,mkfnet,wang2021single,syntanei,wang2019segmenting,tang20203d} has recently been proposed to demonstrate their advances in neuron segmentation studies.
Since the neuronal structures range from long tree-like branches to blob-shape somas, \cite{hanchuan} adopted inception networks~\cite{inception} with various kernel sizes to better learn neuron representations from an enlarged receptive field.
The invention of U-Net~\cite{unet} gave rise to a line of encoder-decoder architectures and popularised them to be one of the de-facto structures in medical image segmentation tasks. 
Under the unsupervised setting, traditional tracing algorithm is combined with 3D U-Net~\cite{3dunet} to progressively learn representative feature and the learned network, in turn, helps improve the tracing performance~\cite{jiezhao}. In the fully-supervised manner, modifications have been made based on 3D U-Net to extend the receptive field of kernels. In MKF-Net~\cite{mkfnet}, a multi-scale spatial fusion convolutional block, where kernels with different size are processed in parallel and fused together, was proposed to replace some of the encoder blocks of 3D U-Net and achieved better segmentation results. Further, \cite{wang2021single} introduces graph-based reasoning to learn longer-range connection for more complete features. The bottom layer of 3D U-Net encodes the richest semantic information but loses the spatial information. To alleviate the loss of spatial cues, \cite{syntanei} proposed to replace the bottom layer with a combination of dilated convolutions~\cite{deeplab} and spatial pyramid pooling operations~\cite{spp}. Although larger local regions have been aggregated after introducing these modifications and better segmentation results have been gained, additional overhead for the inference has also been introduced when using these proposed models. To avoid such overhead, \cite{wang2019segmenting} and \cite{tang20203d} use additional teacher-student model and GAN-based data augmentation techniques, respectively. However, the focus of these segmentation models resides in the learning of local structural information of neurons within a single volume while the intrinsic semantic information of voxels among different volumes has been rarely touched. 

As the semantic label is able to help obtain a better representation for image pattern recognition~\cite{wei2020can,contrastiveseg}, such intrinsic features are beneficial to visual analysis. Siamese networks~\cite{siamese} based unsupervised representation learning methods~\cite{simclr,byol,simsiamese} have recently gained impressive performance over image-level classification task by generating better latent representations through the comparison between two augmented views of the same image. Inspired by these work, we propose to encode the intrinsic semantic features into the representation of each voxel by maximising the similarity between two voxels belonging to the same class in a high-dimensional latent space. In our work, we follow the prevalent encoder-decoder architecture as the base segmentation model. Without the need of negative pairs~\cite{simclr} and momentum encoder~\cite{byol}, we design a class-aware voxel-wise Simple Siamese (SimSiam)~\cite{simsiamese} learning paradigm in a fully-supervised manner to maximise the similarity between two voxels with the same semantic meaning and encourage the base encoder to learn a better latent space for voxels of neuron images. Rather than being restricted within the same volume, to fully utilise the dataset, the voxel pairs can be sampled among different volumes. After training, only the original segmentation base model will be kept. Therefore, no extra cost is required to perform inference. Experimental results on 42 3D optical microscopy neuron images from the BigNeuron project~\cite{bigneuron} show that our proposed framework is able to achieve better segmentation results and further increase the accuracy of the tracing results.

\begin{figure*}[!htb]
\centering     
{\includegraphics[width=\textwidth]{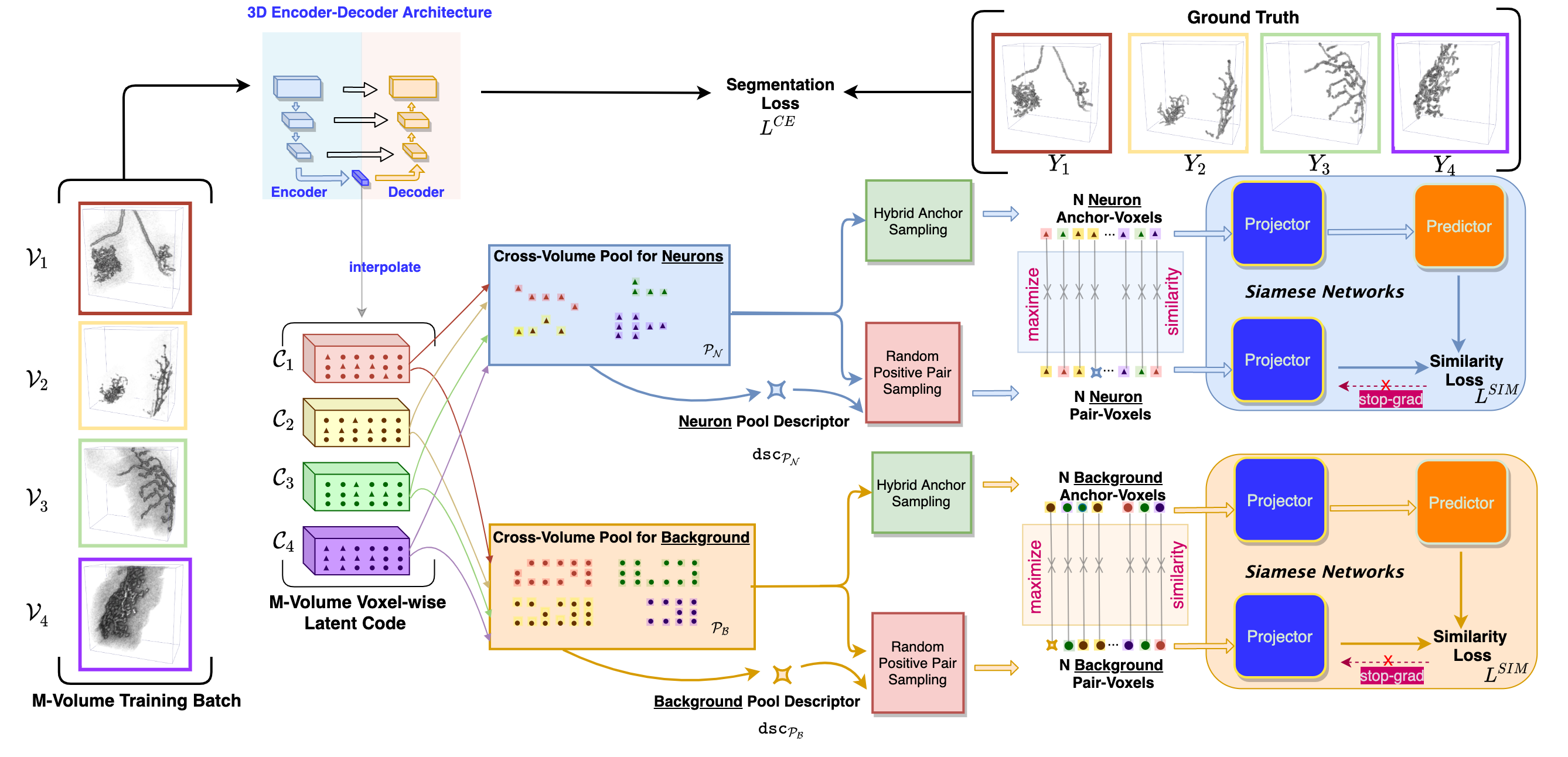}}
\caption{Our proposed training paradigm.} 
\label{fig:method:overview}
\end{figure*}

\section{Method}

\subsection{Supervised Encoder-Decoder Neuron Segmentation}
We apply a 3D encoder-decoder architecture as the base model to perform neuron segmentation. As shown in Fig.~\ref{fig:method:overview}, it consists of an encoder path, a decoder path, and skip connections linking in between, whose training progress is supervised via a binary cross-entropy loss that $L^{CE} = -log\left[Ylog(\widetilde{Y}) + (1-Y)log(\widetilde{Y})\right]$, where $Y$ and $\widetilde{Y}$ represent the ground truth and the predictions of neuron segmentation masks, respectively.

$L^{CE}$ provides semantic cues for each individual voxel but the relation learnt through the encoder is only within a local neighbourhood of each voxel, which ignores the correlation among voxels belonging to the same semantic class in a long distance, i.e., cross-volume. To decrease the distance of voxels with same category in the latent space, we introduce our proposed voxel-wise cross-volume SimSiam representation learning in next section.

\subsection{VCV-RL: Voxel-wise Cross-Volume SimSiam Representation Learning}
As demonstrated in Fig.~\ref{fig:method:overview}, for each volume $\mathcal{V}_i \in R^{D\times H\times W}$ where $D$, $H$, and $W$ denote the depth, height, and width, respectively, we first obtain its corresponding downsized $d$-dimensional latent code through encoders, and then interpolate it back to same size as $\mathcal{V}_i$ to reach a voxel-wise embedding $\mathcal{C}_i$. 
Then, these $M$ volumes of latent codes $\mathcal{C}\in  \mathbb{R}^{M\times D\times H\times W\times d}$ are divided according to the ground truth labels $Y_i$, for the construction of two cross-volume point pools, one for the neurons voxels $\mathcal{P}_\mathcal{N}$ and one for the background voxels $\mathcal{P}_\mathcal{B}$. 
In other words, each pool contains the latent codes of voxels belonging to the same class from all the volumes within a training batch. 

To enforce the voxel embeddings within the same class-aware pool to be closer in their latent space, we adopt Siamese networks~\cite{siamese} to perform the similarity comparison for each input voxel pair, and we denote the voxel to compare as the \textit{anchor-voxel} and the voxel being compared as the \textit{pair-voxel}. Following~\cite{simsiamese}, we use a 3-layer MLP projector $f$ as the channel mapping function to raise the latent code dimension from $d$ to $d_p$, for both types of voxels. Another 2-layer MLP predictor $h$ is employed to match the output of anchor-voxel to that of the pair-voxel. To prevent the Siamese networks from collapsing to constant solution, a stop-gradient operation~\cite{simsiamese} is also adopted as shown in Fig.~\ref{fig:method:overview}.

\subsubsection{Voxel Similarity.} Following SimSiam~\cite{simsiamese}, the symmetrised similarity loss between an anchor-voxel $j$ and a pair-voxel $k$ is defined with \texttt{cos} operator as:
\begin{equation}
    L_{j,k}^{\cos} = \frac{1}{2}(1- \texttt{cos[}h(f(j)), f(k)\texttt{]}),\\
\end{equation}
where $\texttt{cos[}u,v\texttt{]}=\frac{u}{\parallel u\parallel_2}\cdot\frac{v}{\parallel v\parallel_2}$ denotes the cosine similarity between two embeddings. The total similarity loss comparing $N$ voxel pairs is formulated as: 
\begin{equation}
   L^{SIM} = \sum_{*\in[\mathcal{N}, \mathcal{B}]}\frac{1}{N}\sum^{N}_{j\in\mathcal{P}_*, k\in\mathcal{P}_*}(\frac{1}{2}L_{j,k}^{cos}+\frac{1}{2}L_{k,j}^{cos})
\end{equation}

Together with loss $L^{CE}$, the encoded feature embedding for each voxel is expected to contain the intrinsic semantic cues shared among voxels of the same category. As the point-wise cross-entropy loss is complimentary to the representation learning loss~\cite{contrastiveseg}, we keep the weight of these two losses the same. The segmentation loss is then formulated as $L^{seg} = L^{CE}+L^{SIM}$.

\subsubsection{Anchor-Voxel Sampling Strategy.}
Given that latent codes of voxels being misclassified by the base segmentation model are more important for the representation learning~\cite{contrastiveseg}, we design three different strategies to sample $N$ anchor-voxels from each point pool $\mathcal{P}_{*}$:

\begin{itemize}
    \item Random sampling ($\mathcal{AS}_\mathbf{random}$): Randomly sample $N$ anchor-voxels from the whole point pool $\mathcal{P}_{*}$; \item Purely hard sampling ($\mathcal{AS}_\mathbf{PH}$): Randomly sample $N$ anchor-voxels from a subset of point pool $\mathcal{P}_{*}$. The subset is the collection of voxels whose prediction is wrong; 
    \item Hybrid combination sampling ($\mathcal{AS}_\mathbf{hybrid}$): $\frac{N}{2}$ anchor-voxels are randomly sampled from the whole point pool $\mathcal{P}_{*}$ while the rest $\frac{N}{2}$ anchor-voxels are randomly sampled from the subset stated in $\mathcal{AS}_\mathbf{PH}$.
\end{itemize}

\subsubsection{Pair-Voxel Sampling Strategy.} For each selected anchor-voxel, a pair-voxel is sampled from the same point pool to feed together into the Siamese networks. Apart from the candidate points in point pool $\mathcal{P}_{*}$, we propose a virtual point \texttt{dsc}$_{\mathcal{P}_\mathcal{N}}$ and \texttt{dsc}$_{\mathcal{P}_\mathcal{B}}$ as point descriptor for point pool $\mathcal{P}_\mathcal{N}$ and $\mathcal{P}_\mathcal{B}$, respectively to represent an aggregation of semantic feature for each class. We design two ways of computing such a virtual point:

\begin{itemize}
    \item Relaxed: Average pooling of the entire point pool; 
    \item  Strict: Average pooling of the subset of correctly classified points from the entire point pool. 
\end{itemize}

To avoid outliers and stabilise the learning, inspired by Momentum$^2$Teacher~\cite{m2t}, we propose to use a momentum update mechanism to keep the semantic information of past pool descriptors. Formally, the pool descriptor for each pool is defined as:
\begin{equation}
    \texttt{dsc}_{\mathcal{P}_*}^{k} = (1-\alpha)\texttt{dsc}_{\mathcal{P}_*}^{k} + \alpha\texttt{dsc}_{\mathcal{P}_*}^{k-1},
\end{equation}
where $k$ is the current iteration. $\alpha$ is the momentum coefficient and decreases from $\alpha_{base}=1$ to 0 with cosine scheduling policy~\cite{m2t} defined as $\alpha= \alpha_{base} × (\cos(\frac{\pi k}{K}) + 1)/2$, where $K$ is the total number of iterations.

\section{Experiments and Results}
\subsection{Dataset and Implementation Details}
\subsubsection{Dataset.}
Our studies on 3D neuron reconstruction were conducted on the publicly available 42-volume Janelia dataset developed for the BigNeuron project~\cite{bigneuron}, which was further divided into 35, 3, and 4 samples as training, validation, and testing set, respectively. We applied random horizontal and vertical flipping, rotation, and cropping, as data augmentation techniques, to amplify our training set with 3D patches of size $128 \times 128 \times 64$. Given that the number of neuron voxels is dramatically smaller than that of the background voxels, to ensure the existence of foreground neuron voxels, we re-choose the patch until the foreground voxels make up over 0.1\% of the whole patch. 

\subsubsection{Network Setting and Implementation.}
All the models involved in the experiments were implemented in PyTorch 1.5 and trained from scratch for 222 epochs. A model was saved when it reached a better F1-score on the validation set. We use Adam as the optimizer with the learning rate of $1\times 10^{-3}$ and the weight decay of $1\times 10^{-4}$. The batch size $M$ is set as 4. We choose $\mathcal{AS}_\mathbf{hybrid}$ with the number of anchor-voxels $N$ as 512 and strict pool descriptor with MoUpdate mechanism. The hidden layer dimension is 512 and 128 for the projector and predictor, respectively. $d$ and $d_p$ are set as 128 and 512, respectively. To make fair comparison, all the 3D U-Net based segmentation models including Li2019~\cite{syntanei} and MKF-Net~\cite{mkfnet} were implemented with feature dimensions of 16, 32, 64, and 128 for respective layer from top to bottom. As for 3D LinkNet~\cite{linknet}, in addition, we replaced the head and final blocks with convolutional layer without spatial changes to keep the same 3 times downsampling of feature maps.

\subsection{Results and Analysis}
To quantitatively compare among different methods for 3D neuron segmentation, we reported F1, Precision, and Recall as the evaluation metrics. They measure the similarity between the prediction and the ground truth segmentation. Following \cite{syntanei}, we employed three extra metrics for the reconstruction results: entire structure average (ESA), different structure average (DSA), and percentage of different structures (PDS) for the measurement between the traced neuron and the manually annotated neurons.

\subsubsection{Segmentation Results.} The quantitative segmentation result is presented in Table~\ref{tab:segmentation}. Our proposed method achieves the best F1 score among all the other state-of-the-art segmentation methods and improves the performance of the base U-Net by 2.32\%. It is also noticeable that our model includes no additional cost during inference. Our proposed VCV-RL module can be applied on the encoder output of encoder-decoder architecture easily. When applied on the 3D LinkNet, VCV-RL can enhance the segmentation performance by almost 2\%.

\begin{table}[!thb]
\begin{center}
\caption{Segmentation performance on 3D neuron reconstruction.}\label{tab:segmentation}
\begin{tabular}{|l|l|l|l|l|}
\hline
Method &  F1 (\%) & Precision (\%) & Recall (\%) & \texttt{\#Params}\\
\hline
Li2019~\cite{syntanei} & 52.69$_{\pm 12.49}$ & 46.50$_{\pm 12.70}$ & 61.06$_{\pm 11.50}$ & 2.3M\\

3DMKF-Net~\cite{mkfnet}& 52.31$_{\pm 12.14}$ & 46.51$_{\pm 11.82}$ & 59.88$_{\pm 12.32}$ & 1.5M\\
\hline
3D U-Net~\cite{3dunet} & 51.26$_{\pm 12.16}$ & 45.26$_{\pm 11.98}$ & 59.35$_{\pm 12.28}$ & 1.4M\\

\quad \textbf{+ VCV-RL (proposed)}  & \textbf{53.54}$_{\pm 11.39}$ & \textbf{46.68}$_{\pm 11.58}$ & \textbf{63.02}$_{\pm 10.38}$ & \textbf{1.4M} \\
\hline
3D LinkNet~\cite{linknet} & 50.74$_{\pm 12.67}$ & 42.93$_{\pm 11.41}$ & 62.22$_{\pm 14.50}$ & 2.1M\\

\quad + \textbf{VCV-RL}      & 52.66$_{\pm 12.14}$ & 46.10$_{\pm 11.92}$ & 61.55$_{\pm 11.87}$ & 2.1M\\
\hline
\end{tabular}
\end{center}
\end{table}

\begin{table}[!thb]
\begin{center}
\caption{Tracing performance on 3D neuron reconstruction.}\label{tab:tracing}
\begin{tabular}{|l|l|l|l|l|l|l|}
\hline
Method &  ESA $\downarrow$ & DSA $\downarrow$ & PDS $\downarrow$ & F1 (\%)$\uparrow$ & Prec. (\%)$\uparrow$ & Recall (\%)$\uparrow$ \\
\hline
APP2~\cite{app2}                           & 3.62$_{\pm 0.76}$ & 6.80$_{\pm 1.24}$ & 0.34$_{\pm 0.03}$ & 55.84$_{\pm 12.96}$ & 57.48$_{\pm 13.77}$ & \textbf{64.94}$_{\pm 23.10}$\\
+ U-Net~\cite{3dunet}       & 1.59$_{\pm 0.19}$ & 3.66$_{\pm 0.81}$ & 0.22$_{\pm 0.02}$ & 64.92$_{\pm 15.33}$ & 86.23$_{\pm 7.76}$ & 56.30$_{\pm 19.39}$\\
+ MKF-Net \cite{mkfnet}  & 1.62$_{\pm 0.21}$ & 3.85$_{\pm 0.76}$ & 0.22$_{\pm 0.03}$ & 65.15$_{\pm 14.93}$ & \textbf{89.03}$_{\pm 8.74}$ & 55.07$_{\pm 18.13}$\\
+ \textbf{Proposed}        & \textbf{1.52}$_{\pm 0.21}$ & \textbf{3.48}$_{\pm 0.29}$ & \textbf{0.21}$_{\pm 0.02}$ & \textbf{66.17}$_{\pm 14.39}$ & 87.25$_{\pm 7.41}$ & 56.87$_{\pm 17.92}$\\
\hline
\end{tabular}
\end{center}
\end{table}

\begin{figure*}[!htb]
\centering     
\begin{subfigure}{0.17\textwidth}\centering\includegraphics[width=\textwidth]{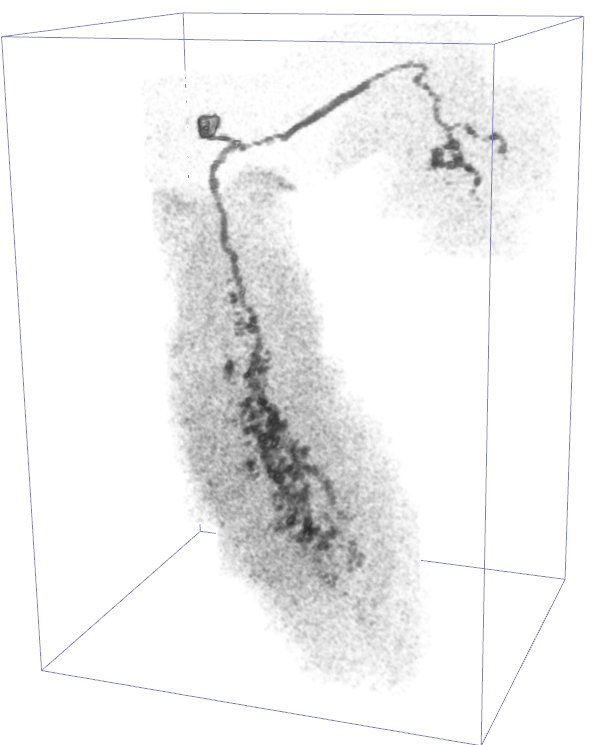}\caption*{\tiny Input}\end{subfigure}%
\begin{subfigure}{0.17\textwidth}\centering\includegraphics[width=\textwidth]{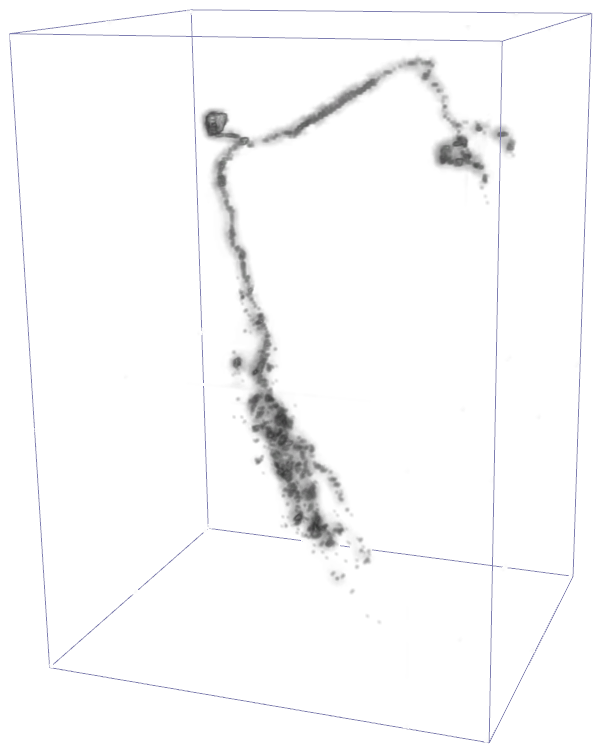}\caption*{\tiny3D U-Net}\end{subfigure}%
\begin{subfigure}{0.17\textwidth}\centering\includegraphics[width=\textwidth]{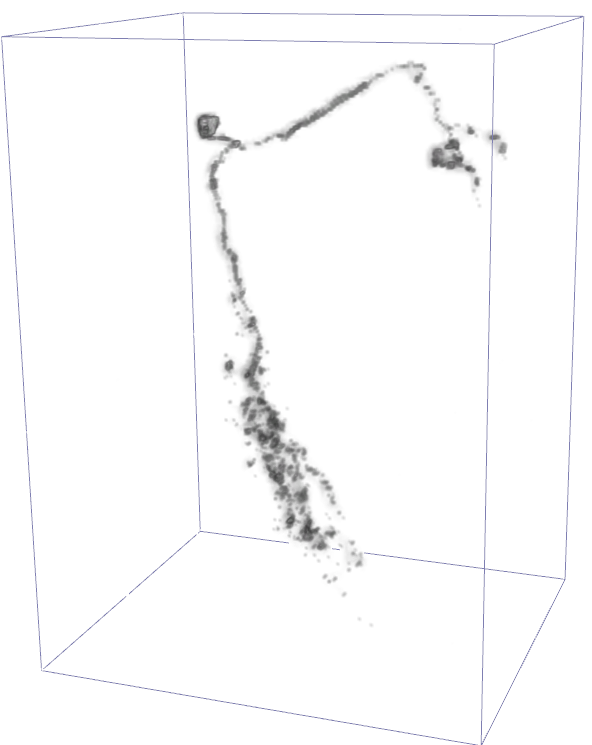}\caption*{\tiny3D MKF-Net}\end{subfigure}%
\begin{subfigure}{0.17\textwidth}\centering\includegraphics[width=\textwidth]{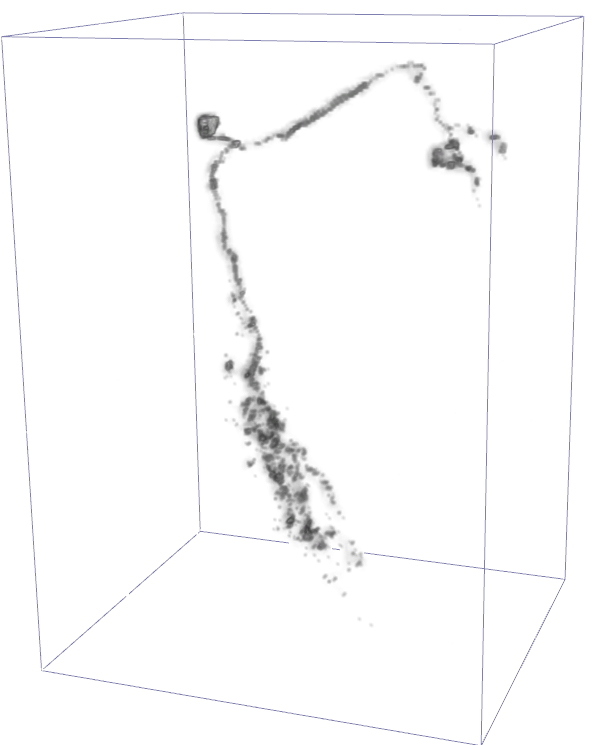}\caption*{\tiny Proposed}\end{subfigure}%
\begin{subfigure}{0.17\textwidth}\centering\includegraphics[width=\textwidth]{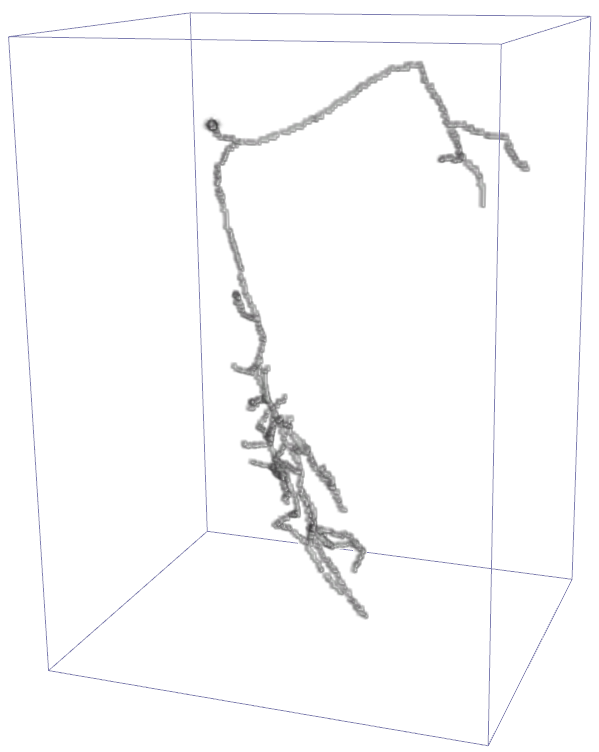}\caption*{\tiny GT}\end{subfigure}\\%
\begin{subfigure}{0.165\textwidth}\centering\includegraphics[width=\textwidth]{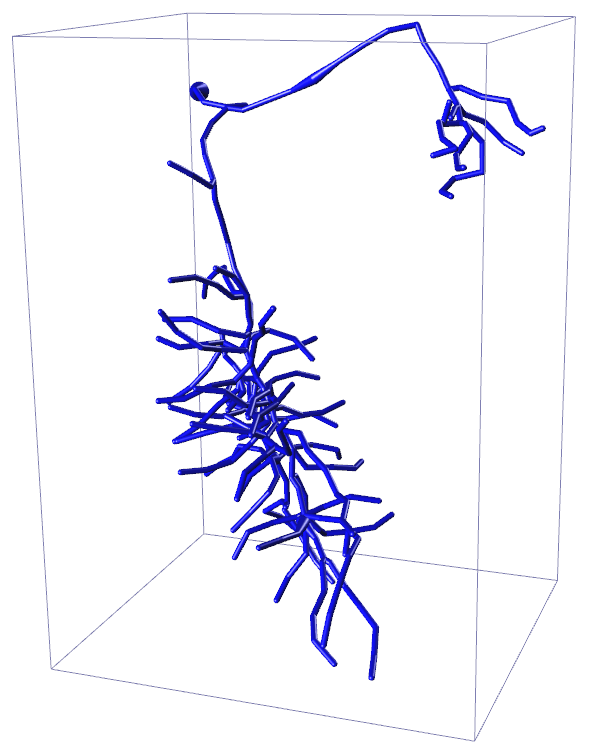}\caption*{\tiny APP2}\end{subfigure}%
\begin{subfigure}{0.165\textwidth}\centering\includegraphics[width=\textwidth]{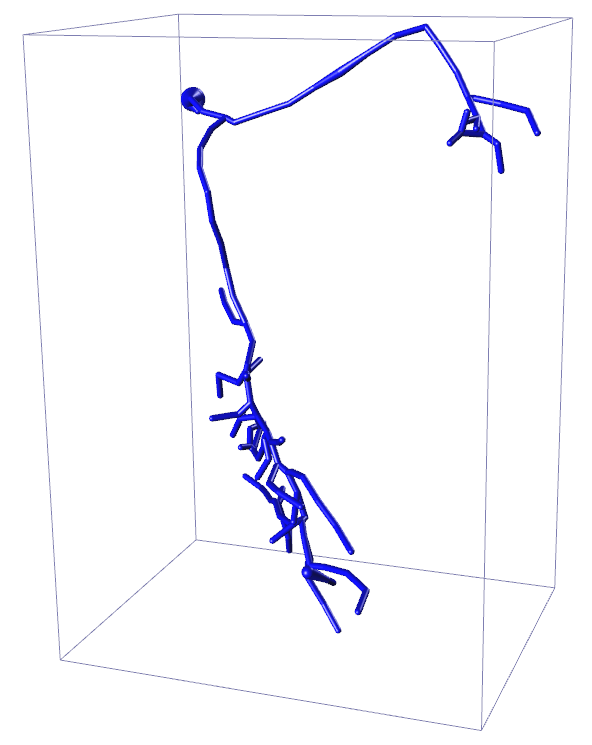}\caption*{\tiny +3D U-Net}\end{subfigure}%
\begin{subfigure}{0.17\textwidth}\centering\includegraphics[width=\textwidth]{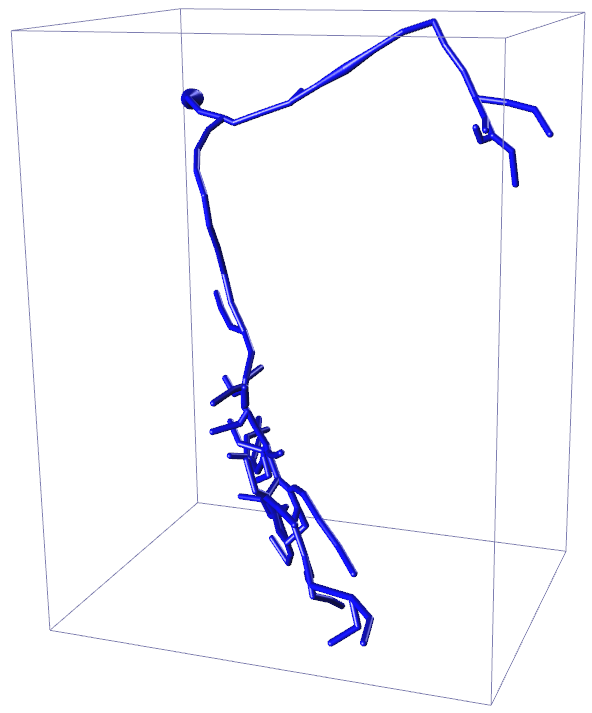}\caption*{\tiny +3D MKF-Net}\end{subfigure}%
\begin{subfigure}{0.16\textwidth}\centering\includegraphics[width=\textwidth]{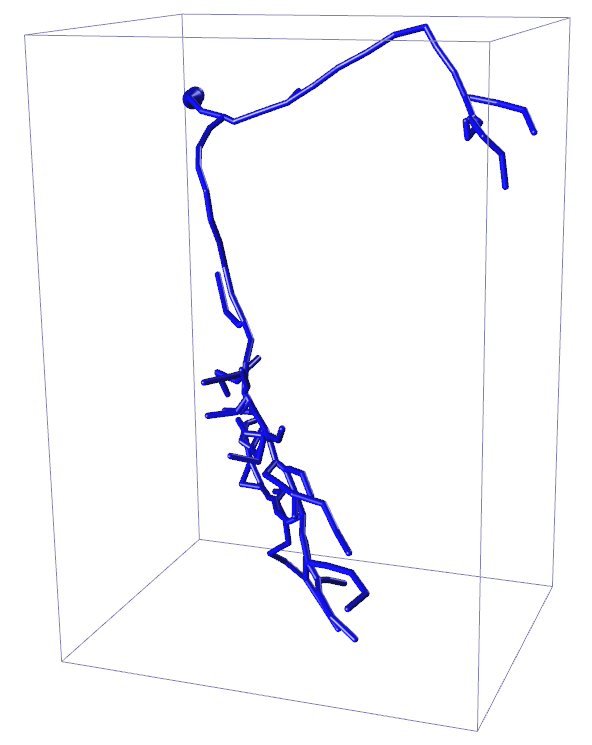}\caption*{\tiny +Proposed}\end{subfigure}%
\begin{subfigure}{0.16\textwidth}\centering\includegraphics[width=\textwidth]{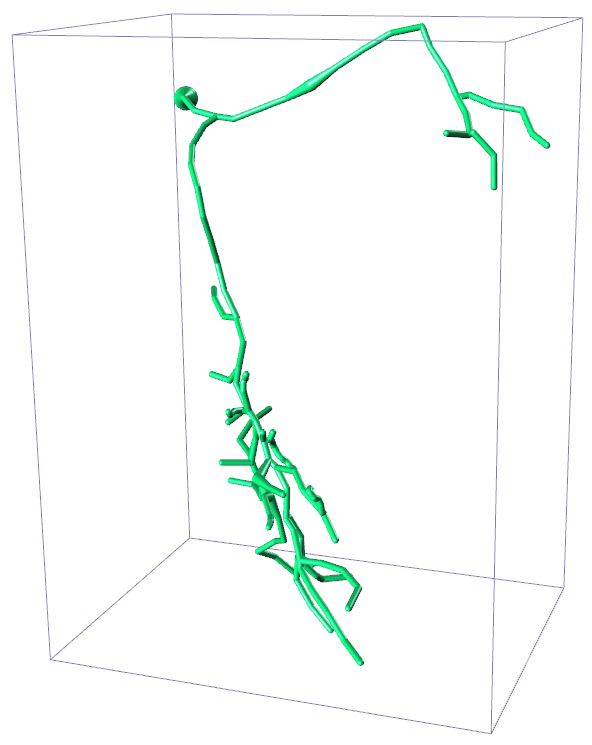}\caption*{\tiny Manual}\end{subfigure}%
\caption{Visualization of segmentation (above) and reconstruction (bottom) results for an example 3D neuron image. $+$ refers performing APP2 on the segmented results from the segmentation model.}
\label{fig:visual_compare}
\end{figure*}

\subsubsection{Neuron Reconstruction Results.} To validate whether the proposed segmentation method can facilitate the tracing algorithm, we choose the state-of-the-art tracing algorithm APP2~\cite{app2} as the main tracer. As in~\cite{hanchuan}, we perform the tracing algorithm on adjusted input volume using the probability map predicted from each segmentation model. As presented in Table~\ref{tab:tracing}, without extra overhead during inference, our proposed method combined with the tracer achieves the best quantitative tracing results on all the metrics among all the other deep learning based reconstruction methods except for the precision. The reason why plain APP2 reaches such high recall is that it overtraces the neuron structures, which is likely to include more real neuron points. Fig.~\ref{fig:visual_compare} displays the enhanced segmentation results after the image adjustment operation proposed in~\cite{hanchuan} in the first row. The second row presents the tracing results after applying APP2~\cite{app2} on the segmented images produced by different segmentation methods.  Our proposed method combined with APP2 achieves competitive tracing result. We note that joint training of segmentation and tracing may further improve the 3D neuron reconstruction performance~\cite{tregidgo20203d}.

\subsubsection{Ablation Study.}
\begin{figure}[!h]
\begin{floatrow}

\capbtabbox{%
  \begin{tabular}{|l|l|l|}
  
    \hline
    ID & Method &  F1 (\%) \\
    \hline
    A & U-Net & 51.26 \\
    B &\textbf{+}$\mathcal{AS}_\mathbf{random}$ & 53.19 \\
    C &\textbf{+}$\mathcal{AS}_\mathbf{PH}$     & 52.69 \\
    D  &\textbf{+}$\mathcal{AS}_\mathbf{hybrid}$ & \\
    D.1 &\quad \textbf{w/o} \texttt{dsc}$_{\mathcal{P}}$         & 53.20\\ 
    D.2 &\quad \textbf{w/} \texttt{dsc}$_{\mathcal{P}}$ (relaxed)  & 53.34\\ 
    \textbf{D.3} &\quad \textbf{w/} \texttt{dsc}$_{\mathcal{P}}$ (strict) & \textbf{53.54} \\ 
    \hline
    E & \textbf{\texttt{D.3}} \textbf{w/o} MoUpdate  & 53.09\\
    \hline
   \end{tabular}
}{%
  \caption{Results of ablation studies.}%
  \label{tab:ablation}
}
\ffigbox{%
  \includegraphics[width=0.5\textwidth]{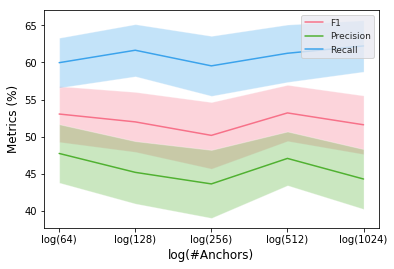}
}{%
  \caption{$\mu \pm \frac{\sigma}{3}$ curves against $\log{N}$.}
  \label{fig:anchor_num_plot}
}
\end{floatrow}
\end{figure}

As presented in Table~\ref{tab:ablation}, model \texttt{B}, \texttt{C}, and \texttt{D.1} reach better F1 score than Model \texttt{A}, which demonstrates the effect of representation learning in improving learning ability of the base U-Net model and the superiority of hybrid anchor sampling strategy. The reason why model \texttt{D.2} and \texttt{D.3} outperform \texttt{D.1} is because the existence of the proposed pool descriptor $\texttt{dsc}_{\mathcal{P}}$ can help improve the generality of the latent space. The strict way of computing $\texttt{dsc}_{\mathcal{P}}$ can further enhance the segmentation performance. We also try to remove the momentum update mechanism of $\texttt{dsc}_{\mathcal{P}}$ and the experiment result of model \texttt{E} demonstrates the importance of past information storage. In addition, we conducted experiments on the proposed method with different number of anchor-voxels $N$. The result is presented in Fig.~\ref{fig:anchor_num_plot}. When $N$ is 512, the F1-score is largest.

\section{Conclusions}
In this paper, we propose a novel voxel-wise cross-volume representation learning with the aim of facilitating the challenging 3D neuron reconstruction task. The segmentation of 3D neuron image is beneficial to improve the performance of tracing algorithms but is challenging due to various imaging artefacts and complex neuron morphology. Recent deep learning based methods rely on specially-designed structures in order to fully utilise the scarce 3D dataset. Though effective, extra computational costs have also been introduced. To make better use of the small dataset without sacrificing the efficiency during inference, we propose a novel training paradigm which consists of a SimSiam representation learning module to explicitly encode the semantic cues into the latent code of each voxel by comparing two voxels belonging to the same semantic category among different volumes. Compared to other methods, our proposed method learns a better latent space for the base model without modifying any part of it. And our proposed method shows superior performance in both the segmentation and reconstruction tasks. 
%
%
%
\bibliographystyle{splncs04}
\bibliography{refs}

\end{document}